**Social and environmental prospects for commercial rangelands: A modelling assessment**

Javier Ibáñez[a] (Corresponding Author), Jaime Martínez-Valderrama[b], Joaquín Francisco Lavado Contador[c], Susanne Schnabel[c], Manuel Pulido Fernández[c]

[a] Departamento de Economía Agraria, Estadística y Gestión de Empresas
ETSIAAB, Universidad Politécnica de Madrid
Ciudad Universitaria s/n, 28040 Madrid
javier.ibanez@upm.es

[b] Estación Experimental de Zonas Áridas
Consejo Superior de Investigaciones Científicas
Almería, Spain

[c] Grupo de Investigación GeoAmbiental
Universidad de Extremadura
Cáceres, Spain

# Social and environmental prospects for commercial rangelands: A modelling assessment

**Abstract**

This paper presents an integrated modelling assessment that estimated the sensitivities of five endogenous factors in commercial rangelands, i.e. number of active farmers, profits, stocking rate, standing herbage biomass, and soil erosion, to the same percentage variation in 70 factors, including economic and climate drivers. The assessment utilised a system dynamics model (107 equations) which represents an area of extensive private farms, its farmers, the main local markets on which they trade, and key ecosystem services involved. The assessment procedure consisted in analysing the behaviours of 288,000 variants of this system during 300 years, each under a different economic and climate scenario.

Our key findings were as follows: 1) It is likely that at least annual grasslands will suffer environmental degradation in the future, and that such degradation will be primarily caused by climate change, not by the increasing demand for livestock products; 2) Private farming systems provide social and economic security to farmers against the effects of climate change, especially in a scenario of rising prices of animal products. However, this research will remain incomplete until its methods and results can be contrasted with other similar assessments.

## 1. Introduction

Rangelands can be defined in different ways (Sayre et al., 2017; Herrick et al., 2012). However, the definition given by Pratt et al. (1966), namely "land carrying natural or semi-natural vegetation which provides a habitat suitable for herds of wild or domestic ungulates", is good enough for the purposes of this paper. Rangelands contribute to the livelihoods of hundreds of millions of people (between 0.7 and 2 billion according to Engler and von Wehrden (2018)), and provide ecosystem services such as forage, fresh water, medicinal products, nutrient cycling, biodiversity, carbon sequestration, hunting, traditional lifestyles, or tourist experiences (Sala et al., 2017). Estimates of the total global area of rangelands vary widely (Reid et al., 2014), not always due to differences in its definition. Thus, for example, Asner et al. (2004) state that managed grazing occupies 25% of the global land surface whereas Engler et al. (2018) put this figure at 44.1%. In any case, rangelands are the form of land use with the largest geographic extent (Ferner et al., 2018; Reid et al., 2014; Asner et al., 2004). Most of the world's grazing area (78% according to Asner et al. (2004)) is in drylands, i.e. areas where precipitation is typically unpredictable and the ratio of mean annual precipitation to mean annual potential evapotranspiration is less than 0.65, as defined by the U.N. Convention to Combat Desertification (Millennium Ecosystem Assessment, 2005).

Rangelands are being affected by several global drivers of change, namely climate change, the increasing demand for food and renewable energy, technological change and the mobility of global capital. Climate change is expected to cause a decrease in precipitation, and increases in potential evapotranspiration, the variability of rainfall and the frequency and duration of droughts in most rangelands. Thus, climate change would aggravate the aridity of these areas worldwide (Reid et al., 2014; Maestre et al., 2012). The increasing demand for livestock products, which is caused by a growing human population and by a shift in diets, especially in the developing world, pushes the prices of animal products up (Reid et al., 2014; FAO, 2012). The increasing demand for cereals, fibre and renewable energy, along with technological change and the mobility of global capital, are already causing a "revolutionary" conversion of rangelands to other land uses, especially crop production (Reid et al., 2014; Herrick et al., 2012).

It is difficult to assess the likely social and environmental impacts of these global drivers on rangelands. Higher prices of animal products would increase the generally low incomes of farmers and pastoralists. But greater aridity would have the contrary effect, because it would reduce the amount of grazing resources and make their provision more uncertain. The outcome of these opposite forces will likely determine the extent to which people, especially young people, continue leaving rangelands (Reid et al., 2014), and the rate of conversion to other land uses. Producers with higher incomes would be encouraged to intensify production, particularly by increasing livestock numbers and by supplementary feeding (Reid et al., 2014). What environmental impacts would such intensification have on rangelands? Many authors state that it would increase the risk of land degradation (Engler et al., 2018; Sayre et al., 2017; Schulze et al., 2016; von Wehrden et al., 2012; Stafford Smith et al., 2007; Vetter, 2005; Illius and O'Connor, 1999), but others disagree (Sullivan and Rohde, 2002; Perevolotski and Seligman, 1998). In any case, defining degradation in a variable system is challenging (Gillson and Hoffman, 2007), and highly subjective (Engler and von Wehrden, 2018; von Wehrden et al., 2012; Reynolds and Stafford Smith, 2002; Perevolotski and Seligman, 1998). On the other hand, the possibility of changing the land use offers new opportunities to rangeland inhabitants and prospects of significant short-term economic benefits, but the environmental effects of such a conversion could be catastrophic if it is carried out in lands that are susceptible to losing soil through erosion (Herrick et al., 2012).

Clearly, the impacts of global drivers on rangelands will depend on what are the characteristics of the land, the temporal and spatial scales, and the groups of interest considered (Reynolds and Stafford Smith, 2002). Thus, the question arises whether it is possible to gain general insights into such impacts. In this paper, we present a study that seems to give an affirmative answer to this question, and which was based on the following three premises:

i. System sensitivities hold crucial clues. For example, one of our estimates is that soil erosion is around eight hundred times more sensitive to an increase in rainfall variability than to an increase, by the same percent, in the world/regional price of meat (Section 3). Hence, climate change would pose a much greater environmental threat to rangelands than the growing demand for animal products (with the corresponding increase in livestock numbers).

ii. Simulation models are needed. Estimating system sensitivities by performing statistical methods on data collected from a representative sample of sites pose major problems. Indeed, an enormous amount of data is required to take account of the many ecological and socio-economic factors involved, their relationships and variability;

much of these data are unavailable, especially social and economic data; and, after all, global drivers could result in changes that were not observed in the past (Engler and von Wehrden, 2018; Oomen et al., 2016; Tietjen and Jeltsch, 2007; Williams and Albertson, 2006; Vetter, 2005).

iii. An integrated multidisciplinary approach is needed. This has been, and still is, widely advocated in the rangeland literature (Engler and von Wehrden, 2018; Ferner et al., 2018; Berrouet et al., 2018; Herrero and Thornton, 2013; Maestre et al., 2012; Herrick et al., 2012; Dougill et al., 2010; Reynolds et al., 2007; Stafford Smith et al., 2007; Gillson and Hoffman, 2007; Vetter, S., 2005). However, integrated studies are rare, mainly because of a lack of communication across disciplines (Vetter, S., 2005). Integration means to take account of the key exogenous and endogenous factors in a rangeland system, and their relationships. An exogenous factor, i.e. a driver in the sense given here, in accordance with Stafford Smith et al. (2007), influences the rangeland but is not, or is hardly, influenced by it; an endogenous factor is influenced by other endogenous factors and, possibly, by some of the exogenous ones. It follows from these definitions that the stocking rate is never a driver, but an endogenous factor whose determinants could differ between rangeland systems. Thus, at least in commercial rangelands, which would currently amount to 45.5% of the total global area of rangelands (Engler et al., 2018), but which are undergoing major expansion (Sayre et al., 2017; Reid et al., 2014; Dougill et al., 2010; Vetter, 2005; Ellis and Swift, 1988), diverse economic, managerial and climatic factors interact to determine stocking rates. Among the economic and managerial factors there are drivers such as world/regional prices, and the advent of technologies, production systems, or policy environments (Stafford Smith et al., 2007); and endogenous factors such as local prices, density of farmers/pastoralists, their economic character, i.e. conservative or opportunistic, level of inputs, or profits. Hence, at least in commercial rangelands, an integrated socio-environmental impact assessment should lead to comparisons of the sensitivities of key endogenous factors (including stocking rates) to diverse drivers, similar to that given to illustrate our first premise. However, notable exceptions notwithstanding, e.g. Dougill et al. (2010), there is a lack of such a kind of assessments. As we will see, a side effect of this deficiency might be that the undeniable potential for livestock to impact rangelands may receive undue emphasis.

In sum, this paper presents a study aimed at assessing, by means of an integrated model, the sensitivities of key endogenous factors in private commercial farms, i.e. number of active farmers, profits, stocking rate, standing herbage biomass, and soil erosion, to 70 biophysical, economic and behavioural factors, including specific characteristics of global drivers of change, e.g. level of and variability in world/regional prices, depth of and variability in rainfall and potential evapotranspiration, or severity and return period of droughts. The paper is structured as follows. Section 2 provides outlines of the model and the assessment procedure. Results are presented in Section 3, and discussed in Section 4. Finally, the main conclusions drawn from the study are summarized in Section 5. Because of space constraints, a detailed description of the model, the complete list of model parameters, and details about the assessment procedure are provided in a Supplementary Document.

## 2. Material and methods

*2.1. Outline of the model*

The model represents a commercial farming area, its farmers, the main local markets on which they trade, and key ecosystem services involved. Climate is characterised by the alternation of dry and wet seasons as well as the regular occurrence of droughts. Herbage mass is made up only of annual species (woody vegetation is periodically removed by ploughing or fire). Farms are extensive private farms which produce weaned animals for sale (hereafter, the output), and farmers normally buy and provide supplementary feed for livestock. Herd/flock sizes are mostly changed by trading with breeding females (young females only replace old females that leave production). Farmers receive subsidies which do not depend on stocking rates.

Although this is not an unreal grazing system, it would be difficult to estimate the fraction of the global area of rangelands that could be characterized in a similar way. Nevertheless, two reasons make this system interesting for our assessment. First, it represents the final stages of some ongoing transformations in rangelands. Indeed, as already indicated, there is a trend worldwide towards converting communal rangelands into extensive private farms (see references above), and the intensification of livestock production favours a replacement of perennial grasses by annuals (Tarrasón et al., 2016; Stafford Smith et al., 2007; Vetter, 2005; Illius and O'Connor, 1999; Milton and Siegfried, 1994). And second, annual grasslands where livestock mobility is limited and animals are supplemented are considered to be grazing systems with the highest risks of degradation (Engler and von Wehrden, 2018; Engler et al., 2018; Sayre et al., 2017; Schulze et al., 2016; von Wehrden et al., 2012; Stafford Smith et al., 2007; Vetter, 2005; Illius and O'Connor, 1999).

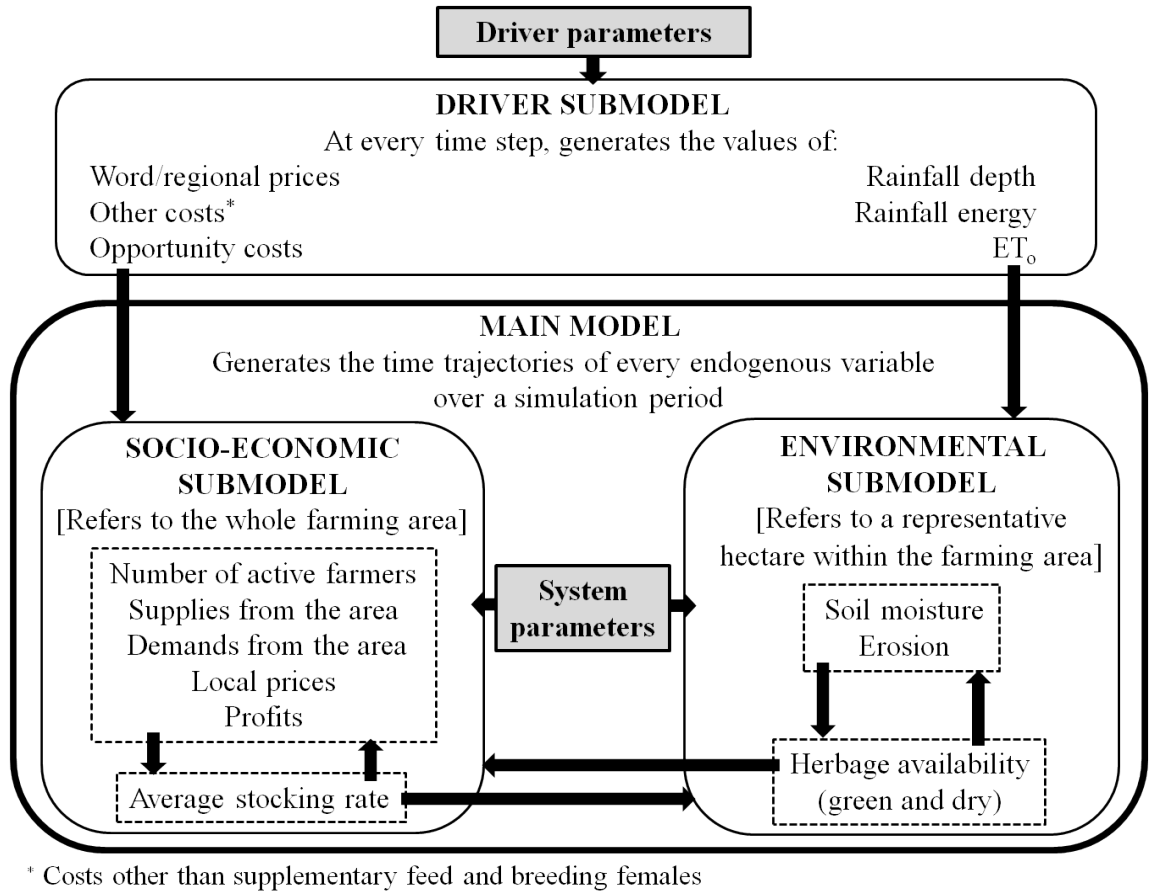

* Costs other than supplementary feed and breeding females

**Fig. 1.** Schematic diagram of the model. Model simulation requires assigning values to the driver and system parameters (in grey).

The model comprises 107 equations which result from the integration and upgrading of models that have already been presented and utilized in previous papers (Ibáñez and Martínez-Valderrama, 2018; Ibañez et al., 2016; Ibañez et al., 2014a). The model follows the system dynamics approach (e.g. Elsawah et al., 2017), and is lumped spatial (Kelly et al., 2013). Therefore, its equations are mostly nonlinear and organized as multiple feedback loops, and the modelled variables are either totals or averages across a certain area, which differ between submodels (see Fig. 1 and below). The model output consists of the time trajectories of every endogenous variable over a predetermined simulation period, and the input is a set of parameter values (Fig. 1). There are 87 biophysical, technical, economic and behavioural parameters, including 11 initial values of stock variables. All of them have real-world counterparts except six calibration parameters (see Table SD1 in the Supplementary Document). Parameters are divided into two groups (Fig. 1):

i. System parameters, which are exclusively involved in the socio-economic and environmental submodels, and whose purpose is to specify the characteristics of the modelled system for a given simulation, i.e. the total area, the potential number of farmers (or farms), livestock characteristics and needs, herbage properties, soil properties, hill-slope length and gradient, subsidies, default sizes of local markets, and average responsiveness of farmers and traders, i.e. the degree at which they are conservative or opportunistic, on average. Besides, system parameters calibrate functional relationships for a given simulation, with each preserving the general non-linear shape (sigmoid, exponential, etc.) that ensures its outcome is not unrealistic even for extreme values of the explanatory variables, and thus model's internal consistency.

ii. Driver parameters, which are exclusively involved in the driver submodel (see below).

Model equations are divided into three submodels (Fig. 1):

i. The driver submodel generates the values of model drivers at every time step. Weather drivers, i.e. rainfall depth, rainfall energy and reference evapotranspiration, are modelled to follow more or less complex random patterns by combining sine wave functions, whose driver parameters are amplitudes and periods, with random distributions, whose driver parameters are means and standard deviations. Economic drivers, i.e. word/regional prices, costs other than supplementary feed and breeding females (hereafter, 'other costs') and opportunity costs of farmers, are modelled to follow interannual cycles by means of sine wave functions. By way of illustration, the generation of rainfall depth during a time step is as follows. Whether it rains or not during a given time step (equal to 0.0078125 years, or around 2.85 days) is the outcome of a Bernoulli random variable in which the 'yes' probability follows seasonal and interannual cycles formulated by means of sine wave functions and specified by means of driver parameters (including the severity and return period of droughts). Then, in case that it rains during the time step, rainfall depth is sampled from a log-normal distribution. This set of equations is a substantial improvement on previous versions.

ii. The socio-economic submodel refers to the whole farming area. It comprises the following endogenous variables and relationships involving system parameters which are not detailed here for the sake of brevity: a) Total number of active farmers (or active farms), which are related to profits and opportunity costs; b) Supplies of output and old females, which are related to the total number of breeding females and amounts of herbage mass and supplementary feed available for livestock; c) Demand for supplementary feed, which is related to the number of animals to be supplemented, availability of herbage, profits and local price of this input; d) External and internal supplies of, and demands for, breeding females, which are related to the number of active farmers, and their entry and exit, profits and local price of breeding females; e) Local prices of output, old animals, supplementary feed, and breeding females, which are related to world/regional prices, supplies/demands from the rangeland, temporary stocks of unsold/undelivered products/inputs, and external demands/supplies, whose behaviour depend on local prices themselves; f) Total profits in the farming area and profits for an average farmer, which are related to the selling of products, the purchasing of inputs, 'other costs', local prices and subsidies; and g) Total number of grazing animals (breeding and young females), which are related to the number of active farmers, profits, and price of breeding females. The average stocking rate results from dividing the total number of grazing animals by the area of active farms.

iii. The environmental submodel refers to a representative hectare within the farming area. It comprises the following endogenous variables and relationships (again, the involved system parameters are not detailed here): a) Soil moisture, which depends on rainfall, evapotranspiration, soil drainage and surface runoff; b) Remaining soil depth, which depends on the weathering rate of the parent rock but mainly on the erosion rate; and c) Green and dry herbage masses, which depend on soil moisture and stocking rate. The erosion rate is related to surface runoff and to the remaining soil depth itself, because bulk density increases as soil is lost. Surface runoff is related to rainfall depth, rainfall energy, soil moisture and canopy cover. Livestock reduce canopy cover by consuming herbage and by creating bare soil areas.

The socio-economic submodel takes the availability of herbage calculated by the environmental submodel as the average availability across the farming area (Fig. 1). Similarly, the average stocking rate across the farming area calculated by the socio-economic submodel is the stocking rate used in the environmental submodel (Fig. 1).

*2.2. Outline of the assessment procedure*

The estimation of system sensitivities was carried out by means of a sensitivity analysis (SA) of the model. This analysis followed a global variance-based method (Gan et al., 2014; Saltelli et al., 2008). The term 'global' indicates that the effect of each parameter $X_i$ (i = 1, ..., k) on a target (endogenous) variable Y is evaluated by varying all parameters in each simulation of the analysis, thereby leading to robust estimates (in contrast to varying one parameter at a time). The term 'variance-based' indicates that the variance of each target variable, V(Y), is decomposed into terms corresponding to the different parameters and their interactions. These terms are commonly summarized by means of two indices: the first order sensitivity index ($S_i$), which measures the first-order effect of $X_i$, i.e. the expected reduction in V(Y) if $X_i$ was fixed; and the total sensitivity index ($ST_i$), which measures the total contribution to V(Y) due to $X_i$, including its first-order effect plus all higher-order effects due to interactions. Clearly, the latter index was especially suitable for our assessment. Given that both indices are always positive, they do not show whether a given effect is positive or negative. When needed, this detail was checked by plotting Y against $X_i$. As estimators of $S_i$ and $ST_i$ (though estimates of the former were dismissed), we used those labelled as $B_3$ by Glen and Isaacs (2012), authors who contrasted the accuracy of 12 pairs of estimators. $B_3$ estimators require simulating the model N(k+2) times, where N is fixed by the user, k is the number of parameters, and 2N is the number of basic scenarios. A basic scenario is a complete set of parameter values (one model input) obtained by randomly and independently sampling a value for each parameter within a predetermined range of variation. To this end, we used Latin Hypercube Sampling. The other kN scenarios were obtained by combining the basic scenarios in the way explained by Glen and Isaacs. In our case, N was equalled to 4000, value at which the estimates of the main sensitivities achieved a satisfactory convergence. A 300-year simulation period was used in order to estimate long-term sensitivities.

We evaluated the sensitivities of five target variables to 70 model parameters. The resting 17 parameters were excluded for different reasons (see the Supplementary Document for details). The target variables were the remaining soil depth at the end of the simulation period [mm], and four averages calculated over the entire simulation period, namely those of the number of active farmers [frm], average stocking rate (across the farming area) [LU/ha], average standing herbage biomass (across the farming area) [kg DM·ha$^{-1}$], and annual earnings per active farmer [€·frm$^{-1}$·yr$^{-1}$] (see the Supplementary Document for details on this variable). All these variables are 'slow' variables, which are useful indicators for gaining insights into long-term changes (Reynolds et al., 2007). The SA utilized 288,000 model runs, i.e. N(k+2) where N = 4000 and k = 70. This means that system sensitivities were estimated by analysing the behaviours of 288,000 variants of the modelled system during 300 years (each one corresponding to a different set of values of the system parameters) under the same number of different external conditions (each one generated on the basis of a different set of values of the driver parameters). Clearly, the power of an analysis like this can only be achieved by means of a simulation model. Note also that running the model under such a large number of scenarios without a single collapse is a strong test of model's internal consistency.

A final point to be stressed is that the estimated sensitivities are normalized, i.e. refer to the same percentage variation in all the parameters. Specifically, each parameter was randomly sampled within a range equal to ±30% of a default value. Most of the default values were derived from data on Spanish dehesas, which are commercial rangelands of the northern Mediterranean region (see the Supplementary Document for details). By setting a wide range around every default value, i.e. by covering a great diversity of variants of the modelled system, we intended to make findings that could be generalized beyond our case study, an issue which is discussed in Section 4.

## 3. Results

Table 1 presents the estimates of $ST_i$ for the 20 most important parameters affecting each target variable of the SA. Regarding these results, we highlight the following points:

i. Socio-economic target variables, i.e. number of active farmers, profits, and stocking rate, were almost exclusively sensitive to socio-economic factors, though they were scarcely sensitive to the price of supplementary feed. Hence, the socio-economic component of private farming systems would be relatively invulnerable to climate change.

Table 1. The rankings of the top 20 most important parameters affecting each target variable, in descending order of $ST_i$. Parameter definitions have been greatly simplified because of space constraints (for full definitions see Table SD1 in the Supplementary Document).

| Avg. Number of active farmers | | Avg. Annual earnings per active farmer | | Av. Stocking rate | | Avg. Standing herbage biomass | | Remaining soil depth | |
|---|---|---|---|---|---|---|---|---|---|
| Parameter | $ST_i$ | Parameter | $ST_i$ | Parameter | $ST_i$ | Parameter | $ST_i$ | Parameter | $ST_i$ |
| Av. WOP[1] level | 0.342 | Av. WOP[1] level | 0.383 | Av. OC[2] level | 0.366 | BD[4] threshold herbage | 0.415 | Initial topsoil porosity | 0.452 |
| Target output per LU | 0.319 | Target output per LU | 0.331 | Av. WOP[1] level | 0.207 | Initial topsoil porosity | 0.397 | BD[4] threshold herbage | 0.328 |
| Av. OC[2] level | 0.197 | Av. OC[2] level | 0.211 | Target output per LU | 0.154 | Av. Rain. depth / TS[8] | 0.166 | SD[10] Rain. depth / TS[8] | 0.260 |
| Potential No. Farmers | 0.132 | Subsidy per farmer | 0.073 | Potential No. Farmers | 0.076 | Av. $ET_o$ per TS[8] | 0.121 | Erosion rate calibration | 0.110 |
| Subsidy per farmer | 0.080 | Energy intake per LU | 0.039 | Energy intake per LU | 0.052 | Av. Rain. Problty. / TS[8] | 0.117 | Av. Rain. Problty. / TS[8] | 0.098 |
| Av. OPC[3] level | 0.072 | Av. OPC[3] level | 0.036 | Subsidy per farmer | 0.052 | GH[9] pot. growth rate | 0.067 | Init. Topsoil erodiblty. | 0.055 |
| Energy intake per LU | 0.032 | Default market output | 0.035 | Default market output | 0.047 | SD[10] Rain. depth / TS[8] | 0.058 | Soil field capacity | 0.045 |
| Period WOP[1] cycles | 0.029 | BD[4] threshold herbage | 0.035 | S. Feed energy content | 0.042 | DH[11] decaying rate | 0.035 | Av. Rain. depth / TS[8] | 0.045 |
| BD[4] threshold herbage | 0.025 | Potential No. Farmers | 0.030 | Amplitude price cycles | 0.037 | SWC[12] field capacity | 0.016 | Random seed | 0.032 |
| S. Feed energy content | 0.023 | S. Feed energy content | 0.030 | BD[4] threshold herbage | 0.037 | $ET_o$ seasonality | 0.015 | Av. Runoff DBS[13] | 0.030 |
| Amplitude price cycles | 0.021 | Period WOP[1] cycles | 0.028 | Snstty. Traders - market | 0.036 | Rainfall seasonality | 0.014 | CV[14] Runoff DBS[13] | 0.027 |
| Initial topsoil porosity | 0.018 | Initial topsoil porosity | 0.025 | Period WOP[1] cycles | 0.028 | Erosion rate calibration | 0.011 | Av. $ET_o$ per TS[8] | 0.024 |
| Snstty. Farmers - prices | 0.018 | Amplitude price cycles | 0.023 | Snstty. Farmers - prices | 0.024 | Init. Topsoil erodiblty. | 0.007 | Rainfall seasonality | 0.020 |
| Snstty. Traders - market | 0.017 | Snstty. Traders - market | 0.019 | Initial topsoil porosity | 0.024 | Runoff calibration | 0.007 | Runoff calibration | 0.012 |
| Av. WSFP[5] level | 0.015 | Snstty. Farmers - prices | 0.016 | Av. WSFP[5] level | 0.019 | Random seed | 0.005 | CV[14] $ET_o$ per TS[8] | 0.012 |
| Default market output | 0.014 | DTTA[6] expectations | 0.015 | DTTA[6] expectations | 0.018 | CV[14] Runoff DBS[13] | 0.004 | DH[11] decaying rate | 0.010 |
| DTTA[6] expectations | 0.013 | Default market S. Feed | 0.015 | Default market S. Feed | 0.018 | Av. OC[2] level | 0.004 | Return period droughts | 0.006 |
| Snstty. Farmers - profit | 0.012 | Av. WSFP[5] level | 0.014 | Period OC[2] cycles | 0.018 | Av. WOP[1] level | 0.002 | $ET_o$ seasonality | 0.005 |
| DTTA[6] targets | 0.009 | Snstvty. Farmers - profit | 0.013 | Av. Rain. depth / TS[8] | 0.013 | Av. Runoff DBS[13] | 0.002 | Canopy cover calibration | 0.004 |
| DTFA[7] expectations | 0.008 | Period OC[2] cycles | 0.009 | DTFA[7] expectations | 0.012 | CV[14] $ET_o$ per TS[8] | 0.002 | GH[9] pot. growth rate | 0.003 |

[1]World/regional output price; [2]Costs other than supplementary feed and breeding females; [3]Opportunity cost; [4]Bulk density; [5]World/regional supplementary feed price; [6]Delay time for traders to adjust ...; [7]Delay time for farmers to adjust ...; [8]Time step; [9]Green herbage; [10]Standard deviation; [11]Dry herbage; [12]Soil water content; [13]Dry and bare soil; [14]Coefficient of variation.

ii. Number of active farmers and profits were mostly sensitive to the same three factors in the same order, namely level of the world/regional price of output (positive relationship), target output per breeding female (positive relationship), and 'other costs' (negative relationship). Subsidies had a much lower effect on both target variables. Hence, unless 'other costs' increase very considerably, it is likely that the growing demand for animal products and the corresponding intensification of livestock production will economically benefit farmers, and will favour their permanence in private farming systems. As stated in point (i), this would be so regardless of climate change. In consequence, it could be taken into consideration the possibility of reducing the level of subsidy.

iii. The stocking rate was mostly sensitive to the same three factors mentioned in (ii), showing the same type of relationships (positive/negative). Hence, again, unless 'other costs' increase significantly, it is likely that the growing demand for animal products and the corresponding intensification of livestock production will increase livestock numbers in private farming systems.

iv. Environmental target variables, i.e. standing herbage biomass and remaining soil depth, were almost exclusively sensitive to biophysical factors. Hence, it is unlikely that the growing demand for animal products and the corresponding intensification of livestock production will impact on forage supply and soil erosion in private farming systems.

v. Environmental target variables were especially sensitive to two parameters: 'initial topsoil porosity' and 'bulk density above which herbage does not grow' (abbreviated as 'BD threshold herbage' in Table 1). These parameters were practically the only biophysical parameters whose effects on the socio-economic target variables, though relatively unimportant, ranked among the top 20 effect list. The importance of both parameters is explained as follows. Given that the parametric scenarios used in the SA were ultimately based on random sampling, in some simulations (specifically, in 83,679 simulations), the value of 'initial topsoil porosity' entailed an initial topsoil bulk density which, by chance, exceeded the value of 'bulk density above which herbage does not grow'. Therefore, in the variants of the modelled system corresponding to these simulations, the soil was entirely denuded of herbage cover from the very beginning. Moreover, it is likely that, in some other simulations (variants), the value of 'initial topsoil porosity' meant that topsoil bulk density was initially below the threshold but sufficiently close to it for the disappearance of herbage mass to happen within the simulation period (recall from Section 2.1 that topsoil bulk density increases as soil is eroded). Hence, the combined effect of the parameters in question can be interpreted as the effect of a shortage of herbage mass. As a corollary, it is likely that shortages of herbage mass caused by a decrease in precipitation will seriously affect soil masses in private farming systems, though will leave their socio-economic components relatively unaffected.

vi. Apart from the parameters mentioned in point (v), standing herbage biomass was mostly sensitive to 'average rainfall depth per time step' (positive relationship), 'average reference evapotranspiration depth per time step' (negative relationship) and 'average probability of rainfall during one time step' (positive relationship). By comparison, parameters related to rainfall variability, including those related to droughts, were relatively unimportant. Hence, it is likely that a decrease in precipitation will significantly reduce the availability of forage, at least in annual grasslands.

vii. Apart from the parameters mentioned in point (v), soil depth was mostly sensitive to rainfall variability, i.e. to 'standard deviation of rainfall depth per time step' (negative relationship). By comparison, parameters related to the amount of rainfall per time step, i.e. 'average probability of rainfall during one time step' and 'average rainfall depth per time step of rain', which showed negative relationships with soil depth, were relatively unimportant. Hence, it is likely that climate change will increase erosion rates, at least in annual grasslands, because the effect of a decrease in precipitation in reducing soil erosion will likely not compensate for the increases caused by a rise in rainfall torrentiality and more exposed soils.

Finally, we should add that the sensitivity of soil depth to the level of the world/regional price of meat was 0.0003 (not included in Table 1). Hence, soil erosion would be around eight hundred times less sensitive to this factor than to rainfall variability, as stated in Section 1.

## 4. Discussion

Overall, the key findings from our modelling assessment are:

Finding #1: It is likely that at least annual grasslands will suffer environmental degradation in the future, and it is likely that such degradation will be primarily caused by climate change, and not by the increasing demand for livestock products.

Finding #2: Private farming systems provide social and economic security to farmers against the effects of climate change, especially in a scenario of rising prices of animal products.

Although the simulation results shown in Section 3 strongly support these findings, they can only be taken as broad statements because modelling invariably involves making simplifying assumptions and dealing with unknowns. Therefore, it is recognized that increases in livestock numbers could significantly contribute to land degradation in certain areas or periods of time, and that other types of grazing systems, e.g. communal systems, could also cope with climate change.

We have not found other studies that support or contradict Finding #1. Many have evaluated the likely effects of climate change and stocking rates on the provision of ecosystem services in rangelands (see references in Section 1). But it seems that no previous study has compared the environmental effects of climate change with those of changes in the drivers of stocking rates, even though this is an essential assessment, because the stocking rate is an endogenous factor, not a driver (Section 1). Our contribution consisted in estimating the sensitivities of five key endogenous factors in commercial rangelands, including stocking rate, to the same percentage variation in 70 factors, including economic and climate drivers. In doing so, each factor could be ranked and compared in importance with others. However, it was beyond the scope of this study to review long-term projections on economic and climate variables. Thus, the question of whether the percentage variations in the former might exceed in the future the percentage variations in the latter to the point that grazing effects will keep up with, or exceed, climate change effects remains unanswered. All we can say here is that it is unlikely (Finding #1), because the estimated sensitivities of the endogenous environmental factors to climate drivers turned out to be hundreds of times greater than those to the economic ones.

Finding #2 may come as no surprise, if only because supplementary feeding is a common strategy for coping with weather variability (Martínez-Valderrama et al., 2018; Müller et al., 2015; Kachergis et al., 2014). On the disadvantage side, Finding #2 implies that farmers would not perceive degradation as a concern, whatever its cause may be. This can be a serious problem affecting commercial agriculture in general, as has been pointed out by some authors (Ibáñez et al., 2014b; Vetter, 2005). However, noting that the model takes for granted that farmers respond exclusively to economic indicators, our assessment shows that, in general terms, such a behaviour would not pose a major threat to the provision of ecosystem services in private farming systems (Finding #1).

It should be clear that our study did not aim at comparing the advantages and disadvantages of different grazing systems, simply because just a single system was evaluated. Therefore, Finding #2 could only be an argument supporting the conversion of communal rangelands into private farms if there is convincing evidence that the former system is vulnerable to climate change. And to find such evidence, more integrated assessments are necessary. In the only study of this type that we know of, Dougill et al. (2010), by also making use of an integrated dynamic model, found that communal systems in the Kalahari region are much more vulnerable to climate change than private farms, even under a market growth scenario, and suggest land privatization as well as a greater sharing of knowledge from private to communal producers as solutions.

Our assessment relies on the adequacies of a model and an assessment procedure. Thus, it is worth considering their strengths and weaknesses, and the extent to which the latter could affect our key findings. Let's start by saying that the model exclusively includes processes and relationships that are widely accepted (well documented) or grounded on a compelling logic. This is why issues such as the possible effects of livestock grazing on water use efficiency and species composition of herbage mass were disregarded. These are highly controversial issues for which, today, there is no widely accepted explanation (Ferner et al., 2018; Sayre et al., 2017; Oñatibia and Aguiar, 2016; Reid et al., 2014; Vetter, 2005; Sullivan and Rohde, 2002; Illius and O'Connor, 1999; Perevolotski and Seligman, 1998; Milchunas and Lauenroth, 1993; Wilson and MacLeod, 1991). Nevertheless, with hindsight, simulation results suggest that both issues would be relatively unimportant when considered within a wide context. Indeed, note that the model formulates the herbage growth rate (out of bare soil areas) as a fraction of the parameter 'potential growth rate of green herbage'. Such a fraction varies with soil moisture over the simulation period. Thus, it equals one, i.e. the actual herbage growth rate equals the potential rate, only if soil moisture is at field capacity and no soil, i.e. no water storage capacity, has been lost by erosion. Otherwise, the fraction is less than one, reaching zero whenever soil water content is at wilting point or when soil mass runs out (see Eq. 29 in the Supplementary Document for details). Therefore, for the model to reflect that livestock affect water use efficiency or cause an undesirable change in herbage composition, the fraction in question should be related to the historical stocking rate, apart from soil moisture, e.g. the additional condition for it to be one would be that the whole area is ungrazed. But the parameter 'potential growth rate of green herbage' (abbreviated as 'GH pot. growth rate' in Table 1) had a negligible effect in all target variables except, needless to say, on standing herbage biomass. Hence, incorporating the mentioned relationship could hardly change the key findings of our assessment. The same reasoning would apply then to other factors affecting herbage productivity such as increased $CO_2$ levels (Maestre et al., 2012; Tietjen and Jeltsch, 2007), the loss of soil nutrients caused by erosion (Illius and O'Connor, 1999) or soil compaction (Pulido et al., 2016). Soil compaction also increases runoff, resulting in higher erosion rates (Asner, 2004). But it is unlikely that incorporating a direct positive relationship between the historical stocking rate and surface runoff could cause socio-economic drivers to enter the list of important effects on soil erosion. This is suggested by the relatively low sensitivity that soil depth showed to the parameters involved in determining surface runoff (Table 1).

The approach adopted for our assessment to consider temporal and spatial variability tried to avoid significantly increasing model complexity, data requirements and unknowns, whereas maintaining the intended generality of the assessment. As explained in Section 2.1, for a given simulation, the model generates temporal variability in drivers, and thus in the entire system, on the basis of a set of parameter values. Also, the representative hectare to which the environmental submodel applies is characterized by means of parameter values for each simulation. Thus, by varying all such values, the SA took account of the long-term evolution of 288,000 hectares differing in the length and gradient of the hill-slope, and in the properties of herbage mass and soil, under the same number of different climate and stocking scenarios. This approach implies sacrificing temporal and spatial precision to generality and realism (Levins, 1966), but it is suitable for a modelling assessment that is aimed at improving understanding of the modelled system and of the way it is expected to react to changes in system drivers (Kelly et al., 2013).

An unavoidable weakness of the model is that many things are assumed to be invariable. Thus, the model does not allow the production system, the policy environment, the technology, or parameter values to change over the simulation period (300 years). However, it does not seem that these simplifications, strong as they may be, can affect the key findings of our study. Indeed, regarding the production system, it can be expected that climate change will encourage farmers to adopt additional strategies for dealing with a variable climate, apart from supplementary feeding. The most salient options are substituting part of the business based on selling weaned animals for a yearling business, diversifying activities on the farm, and using shorter grazing periods (Kachergis et al., 2014; Torell et al., 2010; Müller et al., 2007). But given that these are profitable practices aimed at enhancing the resilience of farming operations and rangeland ecosystems to increasing weather variability, the assumption that farmers will tend to adopt them can only reinforce our key findings.

Concerning the policy environment, it is likely that, under a scenario of a growing demand for livestock products, it will be necessary the incorporation of policies aimed at preventing excessive stocking rates in certain areas or during certain periods of time (recall that Finding #1 is a broad statement). Thus, for example, the European Common Agricultural Policy has already incorporated the cross-compliance system by which payments are conditional upon respecting EU law on environmental, public and animal health, animal welfare and land management (https://ec.europa.eu/agriculture). However, because this kind of policies protect both farmers' income stability and lands, again the assumption that they will be adopted in the future would reinforce our key findings. Similarly, given that technology helps improve labour efficiencies and make better use of inputs, the prospect that new livestock technology will be adopted by farmers would not affect our findings, at least in principle.

Let's consider now the constancy of parameters. We will focus on those parameters that showed important effects on target variables, but are not driver parameters, whose constancy was intended, or initial values, for obvious reasons. These are 'bulk density above which herbage does not grow', 'erosion rate calibration parameter' and 'target for the output per LU' (see Table 1). The first two parameters represent features of the modelled hectare. Certainly, they could change over time in certain circumstances. But taking them as invariable conforms to the intended generality of our assessment. In contrast, it is likely that farmers will aim at maximizing 'target for the output per LU' in the future, because it showed a positive effect on profits. Changing the breed of livestock offers a limited scope for improvement. The most effective alternative would be to incorporate a yearling business, a possibility that has already been commented.

It was explained in Section 2.2 that, although most of the default parameter values used for the assessment were derived from data on a specific case study (Spanish dehesas), we intended to make findings that could be generalized beyond this case. To this end, the parametric scenarios for the SA were generated by sampling each parameter within a wide range around its default value (±30%), i.e. by covering a great diversity of variants of the modelled system and of the external scenarios. This procedure seems acceptable for known or approximated parameters, which are the majority (see Table SD1 in the Supplementary Document), but what about the unknown parameters? It would take long to present a careful examination, so, for the sake of brevity, we will only stress that, fortunately, no unknown parameter ranked among the most important ones, thereby suggesting that our key findings would not be unduly affected by a lack of knowledge.

Our assessment was not aimed at evaluating the impacts of the mobility of global capital and the increasing demand for certain crops and renewable energy. However, it is possible to make two final comments in regard to such drivers. First, it is worth noting that, in the assessment, default values of the managerial parameters (4 parameters) defined farmers as conservative, on average, i.e. as relatively unresponsive to expected profits. More specifically, the default scenario specified that farmers are cautious of changing the size of the reproductive herd or the amount of supplementation provided to animals. In so doing, we just reflected the type of farmers depicted by the interviews with farm owners that provided part of the available data (Ibáñez et al., 2014a). Ibáñez and Martínez-Valderrama (2018), by utilizing a section of the same model used here, found that a widespread conservatism would be optimal for commercial farming areas from economic, social, and ecological perspectives. Thus, it is possible that farmers have managed to find this optimum in areas with a long history of grazing, as is the case of Spanish dehesas. The same authors found that the opposite group behaviour, i.e. a widespread opportunism, would perform better economically, but worse socially and ecologically. It is unlikely that traditional farmers will ever decide to adopt an opportunistic strategy, which is risky and requires special skills and experience to master. However, the mobility of global capital might result in the purchase of rangelands by investors that take livestock as a mere asset, among whom a widespread opportunism could prevail. It is unlikely that our key findings would apply to these rangelands. Second, we have seen that, in general terms, cautiously-managed grazing systems would provide social and economic benefits without reducing the provision of ecosystem services to society. This is a key argument against a careless conversion of rangelands to crop production, a conversion that has often resulted in social crises and environmental collapses (Herrick et al., 2012; Reynolds et al., 2007; Rowntree et al., 2004).

## 5. Conclusions

Rangeland research should address the lack of studies aiming at comparing the social and environmental effects of drivers of change in rangelands, where 'driver' means factor that affects the rangeland but is not affected by it. Clearly, livestock have an undeniable potential for impacting rangelands, and it is crucial to investigate such potential effects. But because the stocking rate is not a driver, it is also crucial to investigate the environmental effects of changes in its determinants. In commercial rangelands, such determinants are economic, climatic and managerial, so integrated assessments, particularly integrated modelling assessments, are required. The need for taking social issues into consideration additionally advocate a multidisciplinary integration.

In this paper, an integrated modelling assessment has been presented. The research has proved very fruitful, and led to clear findings (see Sections 3 and 4). However, it will remain incomplete until its methods and results can be contrasted with other similar assessments

**Acknowledgements**

## References

Asner, G.P., Elmore, A.J., Olander, L.P., Martin, R.E., Harris, A.T., 2004. Grazing systems, ecosystem responses and global change. Annu. Rev. Environ. Resour. 29, 261-299. https://doi.org/10.1146/annurev.energy.29.062403.102142

Berrouet, L.M., Machado, J., Villegas-Palacio, C., 2018. Vulnerability of socio-ecological systems: A conceptual Framework. Ecol. Indic. 84, 632-647. https://doi.org/10.1016/j.ecolind.2017.07.051

Dougill, A. J., Fraser, E. D. G., Reed, M. S., 2010. Anticipating vulnerability to climate change in dryland pastoral systems: Using dynamic systems models for the Kalahari. Ecol. Soc. 15, 14. https://doi.org/10.5751/ES-03336-150217

Ellis, J., Swift, D., 1988. The Dominant Paradigm: Pastoral Degradation of Equilibria1 Ecosystems. J. Range Manag. 41, 450-459. https://doi.org/10.2307/3899515

Elsawah, S., Pierce, S.A., Hamilton, S.H., van Delden, H., Haase, D., Elmahdi, A., Jakeman, A.J., 2017. An overview of the system dynamics process for integrated modelling of socio-ecological systems: Lessons on good modelling practice from five case studies. Environ. Model. Softw. 93, 127145. https://doi.org/10.1016/j.envsoft.2017.03.001

Engler, J.O., von Wehrden, H., 2018. Global assessment of the non-equilibrium theory of rangelands: Revisited and refined. Land Use Policy 70, 479-484. https://doi.org/10.1016/j.landusepol.2017.11.026

Engler, J. O., Abson, D. J., Feller, R., Hanspach, J., von Wehrden, H., 2018. A social-ecological typology of rangelands based on rainfall variability and farming type. J. Arid Environ. 148, 65-73. https://doi.org/10.1016/j.jaridenv.2017.09.009

FAO, 2012. World agriculture towards 2030/2050: The 2012 Revision. ESA Working Paper 12-03, 147. https://doi.org/10.1016/S0264-8377(03)00047-4

Ferner, J., Schmidtlein, S., Guuroh, R. T., Lopatin, J., Linstädter, A., 2018. Disentangling effects of climate and land-use change on West African drylands' forage supply. Glob. Environ. Chang. 53, 24-38. https://doi.org/10.1016/j.gloenvcha.2018.08.007

Gan, Y., Duan, Q., Gong, W., Tong, C., Sun, Y., Chu, W., Ye, A., Miao, C. Di, Z., 2014. A comprehensive evaluation of various sensitivity analysis methods: A case study with a hydrological model. Environ. Model. Softw. 51, 269-285. https://doi.org/10.1016/j.envsoft.2013.09.031

Gillson, L., Timm Hoffman, M., 2007. Rangeland ecology in a changing world. Science, 315, 53-54. https://doi.org/10.1126/science.1136577

Glen, G., Isaacs, K., 2012. Estimating Sobol sensitivity indices using correlations. Environ. Model. Softw. 37, 157-166. https://doi.org/10.1016/j.envsoft.2012.03.014

Guuroh, R.T., Ruppert, J.C., Ferner, J., Čanak, K., Schmidtlein, S., Linstädter, A., 2018. Drivers of forage provision and erosion control in West African savannas-A macroecological perspective. Agric. Ecosyst. Environ. 251, 257-267. https://doi.org/10.1016/j.agee.2017.09.017

Herrero, M., Thornton, P. K., 2013. Livestock and global change: Emerging issues for sustainable food systems. Proc. Natl. Acad. Sci. 110, 20878-20881. https://doi.org/10.1073/pnas.1321844111

Herrick, J.E., Brown, J. R., Bestelmeyer, B. T., Andrews, S. S., Baldi, G., Davies, J., Duniway, M., Havstad, K.M., Karl, J.W., Karlen, D.L., Peters, D.P.C., Quinton, J.N., Riginos, C., Shaver, P.L., Steinaker, D., Twomlow, S., 2012. Revolutionary land use change in the 21st century: Is (Rangeland) science relevant? Rangel. Ecol. Manag. 65, 590-598. https://doi.org/10.2111/REM-D-11-00186.1

Ibáñez, J., Lavado Contador, J.F., Schnabel, S., Pulido Fernández, M., Martínez-Valderrama, J., 2014a. A model-based integrated assessment of land degradation by water erosion in a valuable Spanish rangeland. Environ. Model. Softw. 55, 201-213. https://doi.org/10.1016/j.envsoft.2014.01.026

Ibáñez, J., Martínez-Valderrama, J., Taguas, E. V., Gómez, J. A., 2014b. Long-term implications of water erosion in olive-growing areas in southern Spain arising from a model-based integrated assessment at hillside scale. Agric. Syst. 127, 70-80. https://doi.org/10.1016/j.agsy.2014.01.006

Ibáñez, J., Lavado Contador, J.F., Schnabel, S., Pulido Fernández, M., Martínez-Valderrama, J., 2016. Evaluating the influence of physical, economic and managerial factors on sheet erosion in rangelands of SW Spain by performing a sensitivity analysis on an integrated dynamic model. Sci. Total Environ. 544, 439-449. https://doi.org/10.1016/j.scitotenv.2015.11.128

Ibáñez, J., Martínez-Valderrama, J., 2018. Global effectiveness of group decision-making strategies in coping with forage and price variabilities in commercial rangelands: A modelling assessment. J. Environ. Manage. 217, 531-541. https://doi.org/10.1016/j.jenvman.2018.03.127

Illius, A.W., O'Connor, T.G., 1999. On the relevance of nonequilibrium concepts to arid and semiarid grazing systems. Ecol. Appl. 9, 798-813.

Kachergis, E., J. D. Derner, B. B. Cutts, L. M. Roche, V. T. Eviner, M. N. Lubell, and K. W. Tate., 2014. Increasing flexibility in rangeland management during drought. Ecosphere 5, 1-14. https://doi.org/10.1890/ES13-00402.1

Kelly (Letcher), R.A., Jakeman, A.J., Barreteau, O., Borsuk, Mark E., Elsawah, S., Hamilton, S. H., Henriksen, H.J., Kuikka, S., Maier, H.R., Rizzoli, A.E., van Delden. H., Voinov, A.A., 2013. Selecting among five common modelling approaches for integrated environmental assessment and management. Environ. Model. Softw. 47, 159-181. https://doi.org/10.1016/j.envsoft.2013.05.005

Levins, R., 1966. The strategy of model building in population ecology. Am. Sci. 54, 421-430.

Maestre, F. T., Salguero-Gómez, R., Quero, J. L., 2012. It is getting hotter in here: Determining and projecting the impacts of global environmental change on drylands. Philos. Trans. R. Soc. B Biol. Sci. 367, 3062-3075. https://doi.org/10.1098/rstb.2011.0323

Martínez-Valderrama, J., Ibáñez, J., Del Barrio, G., Alcalá, F.J., Sanjuán, M.E.; Ruiz, A.; Hirche, A.; Puigdefábregas, J., 2018. Doomed to collapse: Why Algerian steppe rangelands are overgrazed and some lessons to help land-use transitions. Sci. Total Environ. 613-614: 1489-1497. https://doi.org/10.1016/j.scitotenv.2017.07.058

Milchunas, D.G., Lauenroth, W.K., 1993. Quantitative effects of grazing in vegetation and soils over a global range of environments. Ecol. Monogr. 63, 327-366.

Millennium Ecosystem Assessment, 2005. Ecosystems and human well-being: Synthesis. Island Press, Washington, DC.

Milton, S. J., Siegfried, W. R., 1994. A Conceptual Model of Arid Rangeland Degradation. BioScience 44, 70-76. https://doi.org/10.2307/1312204

Müller, B., Frank, K., Wissel, C., 2007. Relevance of rest periods in non-equilibrium rangeland systems - A modelling analysis. Agric. Syst. 92, 295-317. https://doi.org/10.1016/j.agsy.2006.03.010

Müller, B., Schulze, J., Kreuer, D., Linstädter, A., Frank, K., 2015. How to avoid unsustainable side effects of managing climate risk in drylands - The supplementary feeding controversy. Agr. Syst. 139, 153-165. https://doi.org/10.1016/j.agsy.2015.07.001

Oesterheld, M., Sala, O.E., McNaughton, S.J., 1992. Effect of animal husbandry on herbivore-carrying capacity at a regional scale. Nature, 356, 234-236.

Oñatibia, G. R., Aguiar, M. R., 2016. Continuous moderate grazing management promotes biomass production in Patagonian arid rangelands. J. Arid Environ. 125, 73-79. https://doi.org/10.1016/j.jaridenv.2015.10.005

Oomen, R.J., Ewert, F., Snyman, H.A., 2016. Modelling rangeland productivity in response to degradation in a semi-arid climate. Ecol. Modell. 322, 54-70. https://doi.org/10.1016/j.ecolmodel.2015.11.001

Perevolotski, A., Seligman, N.G., 1998. Role of grazing in Mediterranean rangeland ecosystems. BioScience 48, 1007-1017.

Pratt, D. J., Greenway, P. J., Gwynne, M. D. 1966. A classification of East African rangeland, with an appendix on terminology. J. Appl. Ecol. 3, 369-382.

Pulido, M., Schnabel, S., Lavado Contador, J.F., Lozano-Parra, J., González, F., 2016. The impact of heavy grazing on soil quality and pasture production in rangelands of SW Spain. Land Degrad. Dev. http://dx.doi.org/10.1002/ldr.2501.

Reid, R. S., Fernández-Giménez, M. E., Galvin, K. A., 2014. Dynamics and resilience of rangelands and pastoral peoples around the globe. Annu. Rev. Environ. Resour. 39, 217-242. https://doi.org/10.1146/annurev-environ-020713-163329

Reynolds, J.F., Stafford Smith, D.M., 2002. Do Humans Cause Deserts? In J.F. Reynolds and D.M. Stafford Smith (eds.), Global Desertification: Do Humans Cause Deserts? Dahlem Workshop Report 88, Dahlem University Press, Berlin, 1-21.

Rowntree, K., Duma, M., Kakembo, V., Thornes, J., 2004. Debunking the myth of overgrazing and soil erosion. Land Degrad. Dev. 15, 203-214. https://doi.org/10.1002/ldr.609

Reynolds, J.F., Stafford Smith, D.M., Lambin, E.F., Turner, B. L., Mortimore, M., Batterbury, S.P.J., Downing, T.E., Dowlatabadi, H., Fernández, R.J., Herrick, J.E., Huber-Sannwald, E., Jiang, H., Leemans, R., Lynam, T., Maestre, F.T., Ayarza, M., Walker, B., 2007. Global Desertification: Building a Science for Dryland Development. Science 316, 847-851.

Sala, O.E., Yahdjian, L., Havstad, K., Aguiar, M.R., 2017. Rangeland Ecosystem Services: Nature's Supply and Humans' Demand. In D.D. Briske (ed.), Rangeland Systems, Springer Series on Environmental Management, DOI 10.1007/978-3-319-46709-2_14, 467-490.

Saltelli, A., Rato, M., Andres, T., Campolongo, F., Cariboni, J., Gatelli, D., Saisana, M., Tarantola, S., 2008. Global sensitivity analysis. The primer. John Wiley & Sons Ltd., Chinchester.

Sayre, N. F., Davis, D. K., Bestelmeyer, B., Williamson, J. C., 2017. Rangelands: Where anthromes meet their limits. Land 6, 31. https://doi.org/10.3390/land6020031

Schulze, J., Frank, K., Müller, B., 2016. Governmental response to climate risk: Model-based assessment of livestock supplementation in drylands. Land Use Policy 54, 47-57. https://doi.org/10.1016/j.landusepol.2016.01.007

Stafford Smith, D.M., McKeon, G.M., Watson, I.W., Henry, B.K., Stone, G.S., Hall, W.B., Howden, S.M., 2007. Learning from episodes of degradation and recovery in variable Australian rangelands. Proc. Natl. Acad. Sci. 104, 20690-20695. https://doi.org/10.1073/pnas.0704837104

Sullivan, S., Rohde, R., 2002. On non-eqilibrium in arid and semi-arid grazing systems. J. Biogeogr. 29, 1595-1618. https://doi.org/10.1046/j.1365-2699.2002.00799.x

Tarrasón, D., Ravera, F., Reed, M. S., Dougill, A. J., Gonzalez, L., 2016. Land degradation assessment through an ecosystem services lens: Integrating knowledge and methods in pastoral semi-arid systems. J. Arid Environ. 124, 205-213. https://doi.org/10.1016/j.jaridenv.2015.08.002

Tietjen, B., Jeltsch, F., 2007. Semi-arid grazing systems and climate change: A survey of present modelling potential and future needs. J. Appl. Ecol. 44, 425-434. https://doi.org/10.1111/j.1365-2664.2007.01280.x

Torell, L.A., Murugan, S., Ramirez, O.A., 2010. Economics of flexible versus conservative stocking strategies to manage climate variability risk. Rangel. Ecol. Manag. 63, 415-425. https://doi.org/10.2111/REM-D-09-00131.1

Vetter, S., 2005. Rangelands at equilibrium and non-equilibrium: Recent developments in the debate. J. Arid Environ. 62, 321-341. https://doi.org/10.1016/j.jaridenv.2004.11.015

von Wehrden, H., Hanspach, J., Kaczensky, P., Fischer, J., Wesche, K., 2012. Global assessment of the non-equilibrium concept in rangelands. Ecol. Appl. 22, 393-399. https://doi.org/10.1016/j.landusepol.2017.11.026

Williams, C., Albertson, J.D., 2006. Dynamical effects of the statistical structure of annual rainfall on dryland vegetation. Glob. Chang. Biol. 12, 777-792, doi: 10.1111/j.1365-2486.2006.01111.x

Wilson, A.D., MacLeod, N.D., 1991. Overgrazing: Present or Absent? J. Range Manag. 44, 475-482. https://doi.org/10.2307/4002748